\documentclass[11pt]{iopart}
\pdfoutput=1
\expandafter\let\csname equation*\endcsname\relax
\expandafter\let\csname endequation*\endcsname\relax
\usepackage[numbers,sort&compress]{natbib} 
\usepackage[pdftex]{graphicx,color}
\usepackage{amsmath,amsfonts,amssymb,wasysym,graphicx,capt-of,ifthen,calc}
\usepackage{enumerate,float,url,color,subfigure}
\usepackage{amsmath}
\usepackage{graphics}  
\usepackage{psfrag} 
\usepackage[latin1]{inputenc}
\usepackage{color}
\usepackage{rotating}
\usepackage{url}
\usepackage{hyperref}
\begin{document}

\title{How Smart Should a Forager Be?}

\author{U. Bhat}
\address{University of California, Santa Cruz, CA 95064}
\author{S. Redner}
\address{Santa Fe Institute, 1399 Hyde Park Road, Santa Fe, NM 87501, USA}

\begin{abstract}
  We introduce an idealized model of an intelligent forager in which higher intelligence corresponds to a larger spatial range over which the forager can detect food. Such a forager diffuses randomly whenever the nearest food is more distant than the forager's detection range, $R$, and moves ballistically towards the nearest food inside its detection range. Concomitantly, the forager's metabolic energy cost per step is an increasing function of its intelligence.  A dumb forager wanders randomly and may miss nearby food, thus making it susceptible to starvation.  Conversely, a too-smart forager incurs a large metabolic cost per step during its search for food and is again susceptible to starvation. We show that the forager's lifetime is maximized at an optimal, intermediate level of intelligence.
  \end{abstract}
\bigskip

\section{Introduction and Model}

Is it possible that nature has optimized the intelligence of foragers? A ``dumb'' forager that possesses no food detection capabilities will forage by random wandering, even if food is nearby.  Such a forager may be at risk for starvation because of its limited ability of find food. Conversely, if a forager is ``smart'' and possesses sensitive food-detection capabilities, it may be able to detect food that is distant. However, the metabolic energy cost of detecting distant food and then traveling to this food source with a body that has an oversized brain may exceed the stored metabolic energy of the forager. Such a forager may also be at risk for starvation.  Here, we address the question of whether there exists an optimal level of intelligence that maximizes the lifetime of a forager. 

To answer this question for realistic forager species and for realistic resource environments is a formidable task that lies beyond the scope of this work. There has been considerable research on various foraging strategies that incorporate some modicum of intelligence, such as, for example, chemotaxis~\cite{purcell1997efficiency,berg1990bacterial}, infotaxis~\cite{vergassola2007infotaxis,martin2010effectiveness}, robotic swarm strategies~\cite{senanayake2016search}, and various types of collective signaling~\cite{traniello1989foraging,tereshko2005collective}.  Our focus is different in that we attempt to address the issue of an optimal intelligence level within the framework of an idealized model of a single forager, namely, the starving random walk model (SRW)~\cite{benichou2014depletion,benichou2016role}. While this naive model has limited realism, we argue that it provides useful insights about the role of intelligence in endowing a forager with a fitness advantage or a fitness detriment. Prior work on the SRW assumed a forager of a fixed size and mobility, and did not incorporate empirically measured parameter values that correlate with the body mass (see, e.g., \cite{kleiber1932body,west1997general}), and was endowed with a nearest-neighbor food detection capability in an extension of the SRW to the "greedy" random walk~\cite{bhat2017does,bhat2017starvation}. In this work, we extend the naive SRW model by using empirically grounded physiological parameters to investigate the role of a variable food detection capability that is coupled to the forager's metabolism on its lifetime.

In the starving random walk model, a forager moves by a nearest-neighbor random walk on a $d$-dimensional lattice with lattice spacing $a$.  This motion naturally describes the motion of a forager without any intelligence. The time for a single step is defined to be $\delta t = {a}/{v}$, where $v$ is the speed at which the forager moves between neighboring sites. After moving one lattice spacing, the forager picks another random direction for its next step; this movement rule corresponds to the forager having a diffusivity $D=a^2/{2\delta t}$. Each lattice site may be empty or contain one morsel of food.  When the forager lands on a food-containing site, the food is entirely consumed and the forager becomes ``full''. This drastic assumption of complete satiation might correspond to a large predator (e.g., a lion) that has killed a large nutritious prey (e.g., a wildebeest). It would be more realistic to posit that the energy content of a single morsel of food is not always sufficient to satiate the forager.  For simplicity, the starving random walk model assumes that the energy content of each food morsel does satiate the forager. Thus after eating, the forager possesses a metabolic energy content of $E_0$, corresponding to satiation.  When the forager takes a single step and lands on an empty site, the forager's metabolic energy content decreases by $dE$. If its metabolic energy $E$ reaches zero, the forager has starved to death. A full forager can thus wander for $N={E_0}/{dE}$ steps until starvation occurs, corresponding to a starvation time of $\mathcal{S}=N\delta t = {a E_0}/{v dE}$.

\begin{figure}[H]
\centerline{\includegraphics[width=0.45\textwidth]{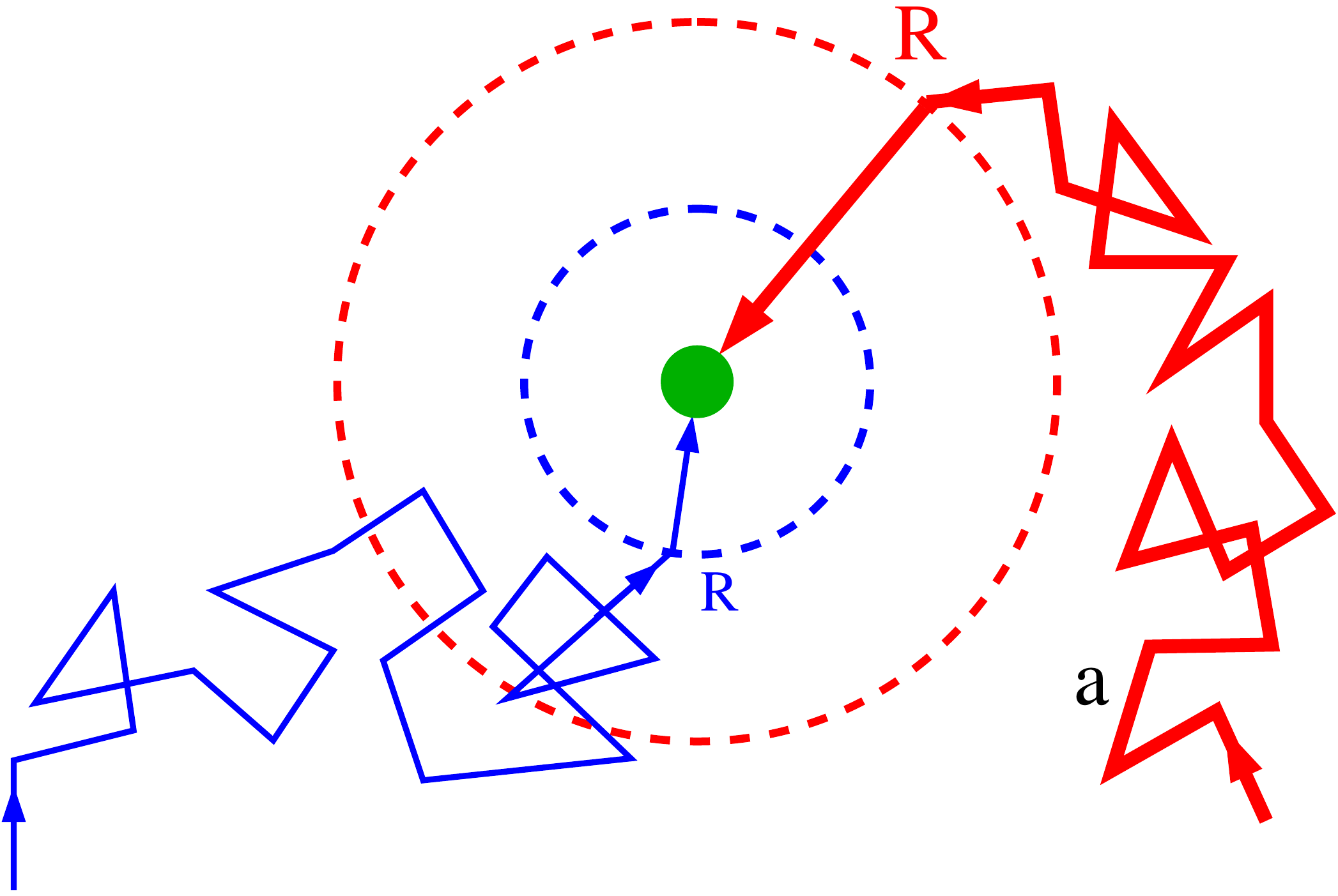}}
\caption{An intelligent forager makes random-walk steps, each of length $a$, when more distant than $R$ from food (green circle).  Once the forager reaches a distance $R$ from the food, it then moves ballistically to the food.  A smart forager (red) has a larger detection radius and thus uses more metabolic energy per step, as indicated by the line thickness, than a dumb forager (blue).}
\label{fig:model}  
\end{figure}

We will investigate the role of intelligence on the forager's lifetime by extending the starving random walk model to endow the forager with a detection range $R$---the distance over which the forager can detect food; a related model was investigated in~\cite{hein2020information}. A larger detection range corresponds to higher intelligence, which, in turn, requires a larger metabolic cost per step on the forager. If the nearest food is a distance $y$ from the forager and $y<R$, the forager moves ballistically towards the food and reaches it in a time $y/v$ (Fig.~\ref{fig:model}).  The energetic cost to ballistically move this distance is $dE\,y/a$.  The constraint that $E_0>dE\,R/a$ must always hold so that a full forager can reach food that is within its detection range. If a forager is more distant than $R$ from food, the forager takes a single random-walk step of length $a$, which again has a metabolic cost $dE$. The food environment is initialized with a Poisson distribution of food morsels with an average spacing $\ell$.  This resource gets depleted by the forager through consumption over the course of its lifetime. These basic parameters are summarized in Table~\ref{tab:parameters}.  For the sake of simplicity, we make the assumption that the metabolic costs of a diffusive step and a ballistic step are equal. If the forager has not starved after a diffusive step, it determines if food is now within the detection range $R$ from its new location.  This foraging continues until either food is found, consumed, and the foraging process begins anew, or the forager starves. 
\begin{table}[ht]
    \centering
        \caption{Basic parameters in the intelligent forager model.}\medskip
    \begin{tabular}{|l|c|}
        \hline
        Quantity & Parameter \\
        \hline
        step size & $a$\\
        time step & $\delta t$ \\
        velocity & $v = a/\delta t$\\
        diffusion coefficient & $D = a^2/2\delta t$\\
        detection range & $R$\\
        metabolism without intelligence& $B$\\
        metabolism with intelligence & $B+\alpha R$\\
        average distance between food  & $\ell$\\
        metabolic energy of a full forager & $E_0$\\
        survival time without food & $\mathcal{S}$\\
        \hline
    \end{tabular}
    \label{tab:parameters}
\end{table}

We assign a metabolic cost for each step of the forager that is an increasing function of the forager's intelligence (see Table~\ref{tab:parameters}).  That is, a forager with a larger detection range is ``smarter'', and the dependence of the energy per step, $dE$, on $R$ is a basic ingredient of our modeling. For simplicity, we posit that $dE=(B+\alpha R)\delta t$, where $B$ is the metabolic energy cost per step for an unintelligent forager and the term $\alpha R$ is the additional metabolic cost per step associated with intelligence.  

Our goal is to determine the lifetime of this forager as a function of its intelligence.  As we will show by an analytical description of the foraging process in one dimension and through simulations in one, two and three dimensions, a forager maximizes its lifetime by possessing an intermediate level of intelligence.  That is, it doesn't pay to be too dumb or too smart.


\section{Preliminary: Starving Random Walk with Sparsely Distributed Food}

As a preliminary, we first determine the average lifetime $\langle T\rangle$ of the starving random walk, that is, a forager with no intelligence, when the food is sparsely distributed.  If every lattice site initially contains food, the average forager lifetime was shown to be $\langle T\rangle \approx 3.2679\,\mathcal{S}$, with the amplitude exactly calculable~\cite{benichou2014depletion}.  For both simplicity and because this approach will be used for the intelligent forager, we now present a heuristic approach that reproduces the correct linear dependence of the lifetime on $\mathcal{S}$, with a nearly correct amplitude. In this approach, the dynamics is partitioned into an early-time and a long-time regime.  At early times, the length of the one-dimensional food desert that is carved out by the forager as it wanders, is sufficiently small that the forager can typically traverse across the desert without starving.  Eventually a critical time is reached where the desert is sufficiently long that a forager typically dies if it attempts to traverse the desert by a random walk.

Beyond this critical time, the far side of the desert becomes irrelevant and we only need consider the forager dynamics in a semi-infinite desert~\cite{bhat2017does,bhat2017starvation}. The time needed to carve out the critical-length desert grows as $\frac{3}{2}\mathcal{S}$~\cite{benichou2016role}, while the survival time of the forager in the semi-infinite desert grows as $2\,\mathcal{S}$.  Accordingly, the average forager lifetime scales as $\frac{7}{2}\,\mathcal{S}$, which is close to the exact result. We now adapt this approach to the situation where the initial food density is small.  This provides the starting point to treat the dynamics of an intelligent forager.  

\begin{figure}[ht]
\centerline{\includegraphics[width=0.5\textwidth]{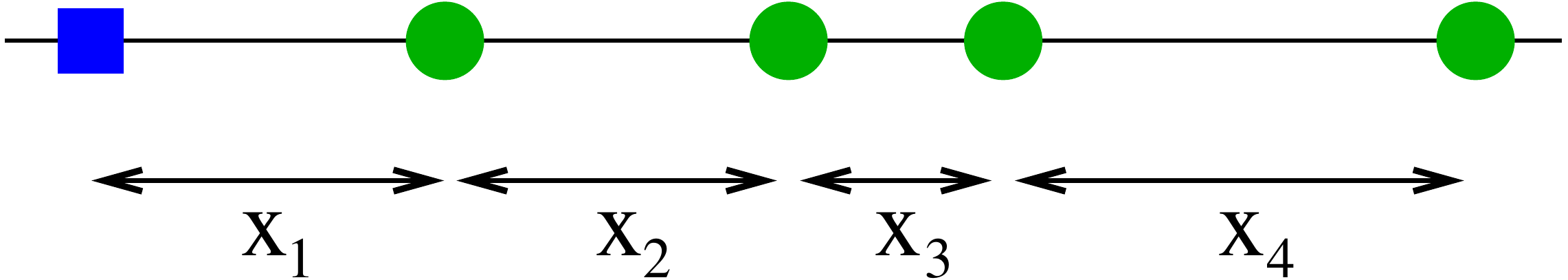}}
\caption{Initial state of the semi-infinite system. The square denotes the forager and circles denote morsels of food.  The region to the left of the forager is empty.}
\label{cartoon}  
\end{figure}

For reasons that will shortly become clear, we treat only the semi-infinite geometry. Suppose that the forager has just eaten, so that it is ``full'', and let $x$ be the distance from its current position to the nearest food. When this distance is larger than the typical range that the forager can diffuse in a time $\mathcal{S}$, $x>\sqrt{D\mathcal{S}}$, the forager is likely to starve without encountering food again. Moreover, if $x$ is of the order of $\sqrt{D\mathcal{S}}$ or larger, the early-time time regime during which a critical-size desert is carved out does not exist because the far side of the desert is effectively unreachable after the first morsel of food is eaten. Thus to estimate the lifetime of the forager, we only need consider the one-sided problem~\cite{bhat2017does,bhat2017starvation}, in which the distance between the forager and successive morsels of food to the right are $x_1,x_2,x_3,\ldots$ (Fig.~\ref{cartoon}).  The lifetime of the forager is the sum of the times to reach each food morsel plus $\mathcal{S}$, the time of the last wandering segment of the forager in which it just starves.

For a full forager that is a distance $x_n$ from the closest food, the probability that the forager reaches it before starving is~\cite{redner2001guide,bray2013persistence,bhat2017does,bhat2017starvation}
\begin{align}
  \label{Hn}
  \mathcal{H}_n = \int_0^{\mathcal{S}} dt\,  F(x_n,t)=
  \text{erfc}\big(x_n/\sqrt{4D\mathcal{S}}\big)\,.
\end{align}
Here $F(x_n,t)$ is the first-passage probability for a diffusing particle to first reach a point that is a distance $x_n$ away at time $t$.  The average time for the excursion to this nearest food is
\begin{align}
  \label{tau-n}
  \tau_n = \frac{\int_0^{\mathcal{S}} dt\, t\, F(x_n,t)}
  {\int_0^{\mathcal{S}} dt\,  F(x_n,t)} =
  \frac{e^{-x_n^2/4D\mathcal{S}}}{\sqrt{\pi}\;{\text{erfc}(x_n/\sqrt{4D\mathcal{S}})}}\, \sqrt{\frac{\mathcal{S} x_n^2}{D}}-\frac{x_n^2}{2D}\,.
\end{align}
The limiting behaviors of this expression are 
\begin{align}
  \label{tau-n-f}
  \tau_n \simeq
  \begin{cases}
{\displaystyle \sqrt{\frac{\mathcal{S} x_n^2}{\pi\,D}} - \frac{x_n^2}{2D}}    &\qquad
{\displaystyle \mathcal{S}\gg \frac{x_n^2}{D}}\,,\\[5mm]
\mathcal{S}    &\qquad {\displaystyle \mathcal{S}\ll \frac{x_n^2}{D}}\,.
\end{cases}    
\end{align}
In the former case, the excursion time is less than $\mathcal{S}$, so that the forager can reach the next morsel of food before starving.  In the latter case, the forager does not reach the next food morsel and thus starves after a time $\mathcal{S}$

Using Eqs.~\eqref{Hn} and \eqref{tau-n-f}, we now write a series representation for the average lifetime $T$ of the forager for a given set of inter-food distances $\{x_n\}$:
\begin{align}
  \label{T-sum}
  T = \tau_1\mathcal{H}_1(1\!-\!\mathcal{H}_2)+ (\tau_1\!+\!\tau_2)\mathcal{H}_1\mathcal{H}_2(1\!-\!\mathcal{H}_3)+
  (\tau_1\!+\!\tau_2\!+\!\tau_3)\mathcal{H}_1\mathcal{H}_2\mathcal{H}_3(1\!-\!\mathcal{H}_4)+\ldots + \mathcal{S}\,.
\end{align}
The first term accounts for the trajectory in which the forager eats the closest food then starves.  The second term accounts for the trajectory in which the forager eats two food morsels and then starves, etc. The last term accounts for the final portion of each trajectory in which the forager does not reach food within a time $\mathcal{S}$ and starves. We now average this series over the distribution of distances $x_n$.  Since these distances are all independent, averages such as $\langle \mathcal{H}_1\mathcal{H}_2\rangle$ and $\langle \tau_1\mathcal{H}_2\rangle$ reduce to $\langle \mathcal{H}\rangle^2$ and $\langle \tau\rangle\langle\mathcal{H}\rangle$, respectively, while products that involve the same index do not decouple.  Thus the average forager lifetime, averaged over all forager trajectories and over all distributions of food is
\begin{align}
  \label{T}
  \langle T\rangle &=\langle \tau \mathcal{H}\rangle\langle 1-\mathcal{H}\rangle
+2\langle \tau \mathcal{H}\rangle \langle \mathcal{H}\rangle\langle 1-\mathcal{H}\rangle
+3\langle\tau \mathcal{H}\rangle \langle \mathcal{H}\rangle^2\langle 1-\mathcal{H}\rangle+\ldots+ \mathcal{S}
  \nonumber\\
&   =  \frac{\langle \tau\mathcal{H}\rangle}{1-\langle\mathcal{H}\rangle}+ \mathcal{S}\nonumber\\
   &= \frac{\langle \tau\mathcal{H}\rangle}{1-\langle\text{erfc}(x/\sqrt{4D\mathcal{S}})\rangle}+ \mathcal{S}\,.
\end{align}
When the food is equally spaced, the above expression reduces to $\langle T\rangle \simeq 2\mathcal{S}$ for $\mathcal{S}\gg x^2/D$ and $\langle T\rangle \simeq \mathcal{S}$ for $\mathcal{S}\ll x^2/D$.  The former limit agrees with previous results~\cite{bhat2017does,bhat2017starvation}, while the latter limiting case corresponds to the trivial situation where the forager fails to find any food and starves without ever eating.

\section{Intelligent Foraging}

We now extend the above results to determine the lifetime of an intelligent forager. To have non-trivial dynamics in which the forager can eat multiple morsels of food before starving, the forager must have sufficient metabolic energy immediately after eating to be able to reach the next morsel of food. This event depends on the relative magnitudes of the basic lengths in the problem---the detection range $R$, the distance to the nearest food, $x$, and the ballistic range of the forager, $v\mathcal{S}$.  The distance that a forager can diffuse before starving, $\sqrt{D\mathcal{S}}$, plays a secondary role, as we discuss below.

\begin{figure}[ht]
    \centerline{
    \includegraphics[width=.9\textwidth]{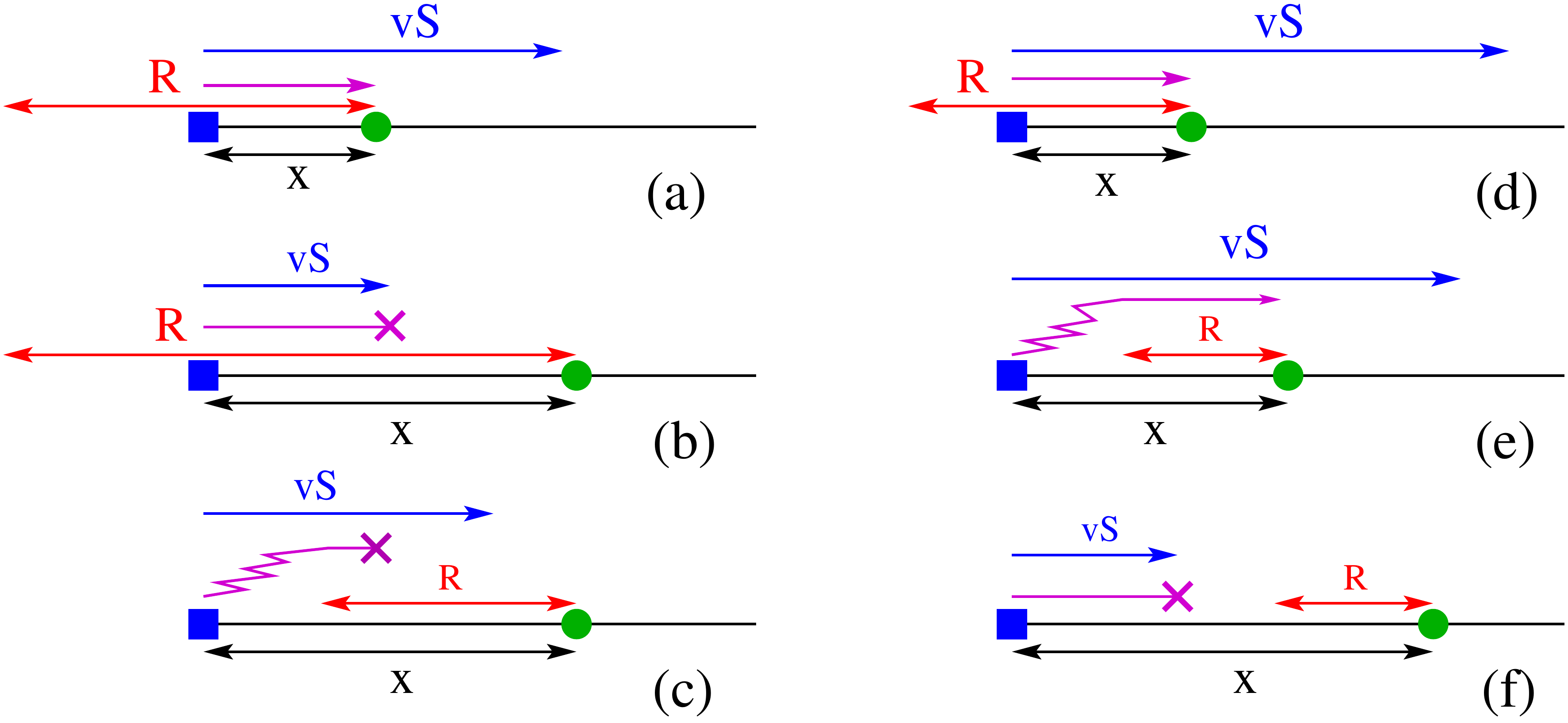}}
    \caption{Configurations that illustrate the distinct inequalities between $R, x$, and $v\mathcal{S}$.  The magenta indicates the forager's trajectory.  In (b), (c), and (f), the $\textcolor{magenta}{\times}$ indicate that the forager (blue square) starves before reaching the food (green dot). In (e), the forager first diffuses (magenta trajectory) to the detection range and may (the case shown) or may not have sufficient metabolic reserve at this point to reach the food.}
    \label{fig:outcomes}
\end{figure}

In terms of $R$, $x$, and $v\mathcal{S}$, it is useful to categorize the possible outcomes of a foraging segment immediately after an intelligent forager has eaten according to whether it is \emph{feeble}, $v\mathcal{S}<R$, or \emph{effective}, $v\mathcal{S}>R$ (Fig.~\ref{fig:outcomes}).  An intelligent, but feeble forager does not possess sufficient metabolic reserve to reach food that is outside its detection range (Fig.~\ref{fig:outcomes}(c)).  Conversely, an intelligent, but effective forager possesses sufficient reserve to potentially reach food that is outside its detection range (Fig.~\ref{fig:outcomes}(e)). For a feeble forager, the possible outcomes are:
\begin{itemize}
    \item[(a)] $x<v\mathcal{S}<R$.  The nearest food is close and the forager necessarily reaches it.
    \item[(b)] $v\mathcal{S}<x<R$ and (c) $v\mathcal{S}<R<x$.  The nearest food is too distant and the forager necessarily starves without eating again.
\end{itemize}
For an effective forager $v\mathcal{S}>R$, the possible outcomes are:
\begin{itemize}
    \item[(d)] $x<R<v\mathcal{S}$. The nearest food is close and the forager necessarily reaches it.
    \item[(e)] $R<x<v\mathcal{S}$.  In this non-trivial case, the nearest food may be reached if the forager diffuses to the detection range $R$ with sufficient remaining metabolic energy to then reach the food by ballistic motion.
    \item[(f)] $R<v\mathcal{S}<x$. The nearest food is too distant and the forager necessarily starves without eating again.
\end{itemize}

To determine the average forager lifetime, let us first focus on the subtlest case (e), specifically $R\lesssim x<v\mathcal{S}$. Starting with the forager at $x=0$ and the nearest food at a distance $x$ away, the probability that the forager reaches the edge of the detection radius at $x-R$ within the requisite time  $\mathcal{S}-R/v$ is (compare with Eq.~\eqref{Hn})
\begin{align}
  \label{Htilde}
  \widetilde{\mathcal{H}}(x,R)=\int_0^{\mathcal{S}-R/v} dt\,  F(x-R,t)=
   \text{erfc}\bigg(\frac{\widetilde{x}}{\sqrt{4D\widetilde{\mathcal{S}}}}\bigg)\,,
\end{align}
where for notational simplicity, we define $\widetilde{x}\equiv x-R$ and $\widetilde{\mathcal{S}}=\mathcal{S}-R/v$.  If the forager reaches $\widetilde{x}$ within a time $\widetilde{\mathcal{S}}$, it necessarily possesses sufficient metabolic reserve to reach the nearest food before starving, as shown in Fig.~\ref{fig:outcomes}(e).  Hence we may again interpret \eqref{Htilde} as the hitting probability for an intelligent forager. Following the same reasoning as that used to obtain Eq.~\eqref{tau-n}, the average time $\tau_C$ of this combined diffusive plus ballistic excursion to reach the nearest food that is a distance $x$ away is
\begin{align}
  \label{tnR}
  \tau_C = \frac{\int_0^{\widetilde{\mathcal{S}}} dt\, t\, F(\widetilde{x},t)}
  {\int_0^{\widetilde{\mathcal{S}}} dt\,  F(\widetilde{x},t)}+\frac{R}{v}\,.
\end{align}
The first term is essentially the same as Eq.~\eqref{tau-n}, except that the forager has to diffuse to within the detection range of the food in a time that is less than or equal to $\widetilde{\mathcal{S}}$.  Once the detection range is reached within this time, the remaining time to reach the food is simply the ballistic transit time $R/v$.  

We now calculate the average forager lifetime by including the contributions from all the configurations in Fig.~\ref{fig:outcomes}. For concreteness and simplicity, we assume that the food is Poisson distributed with average separation $\ell$. Immediately after the forager eats, the probability that the next food lies within a distance $y$ is
\begin{align}
    P(y)= \int_0^y\frac{dx}{\ell}\, e^{-x/\ell} = 1 - e^{-y/\ell}\,.
\end{align}
For a feeble forager, the probability that it reaches the next food before starving is thus $P= 1-e^{-v\mathcal{S}/\ell}$.  This forager therefore consumes 
\begin{align}
    \langle n\rangle =\sum_{n\geq 0} n P^n (1-P)=\frac{P}{1-P}\,
\end{align}
food morsels before it starves.  

The average time that a feeble forager spends in this (successful) ballistic trajectory segment is 
\begin{align}
\label{tauB}
   \tau_B(v\mathcal{S})=  \frac{\displaystyle \int_0^{v\mathcal{S}} \frac{dx}{\ell} \, e^{-x\/\ell}\; \frac{x}{v}}{\displaystyle \int_0^{v\mathcal{S}} \frac{dx}{\ell} \, e^{-x\/\ell}}
 =\frac{\ell}{v}\; \frac{\displaystyle \left[1-e^{-v\mathcal{S}/\ell}\,(1+v\mathcal{S}/\ell)\right]}{\displaystyle (1-e^{-v\mathcal{S}/\ell})}\,.
\end{align}
Thus we estimate the average lifetime $\langle T\rangle$ of a feeble forager to be
\begin{align}
\label{T-poor}
   \langle T\rangle \simeq \frac{P}{1-P} \;\tau_B(v\mathcal{S})+\mathcal{S}\,.
\end{align}
The first term accounts for the $\langle n\rangle$ segments in which the forager reaches food before starving and the trailing term $\mathcal{S}$ accounts for the final segment where the forager starves. 

For an effective forager ($v\mathcal{S}>R$), we need to account for the two cases in which the forager eats, either by a ballistic trajectory to the food or by first diffusing and then moving ballistically to the food (Fig.~\ref{fig:outcomes}(d) \& (e)). The probability for these two events is 
\begin{align}
 Q=\int_0^R \frac{dx}{\ell} e^{-x/\ell} +\int_R^{v\mathcal{S}} \frac{dx}{\ell}\; e^{-x/\ell}\times \widetilde{\mathcal{H}}(x,L)\equiv Q_B+Q_D\,.
\end{align}

The average number of food morsels that the forager consumes before it starves is now $\langle m\rangle =Q/(1-Q)$. We divide these $\langle m \rangle$ segments into $\langle m \rangle (Q_D/Q) = Q_D/(1-Q)$ diffusive segments and $\langle m\rangle (Q_B/Q) = Q_B/(1-Q)$ ballistic segments, with average duration of these two segment types given by Eqs.~\eqref{tnR} and\eqref{tauB}, respectively. We must also include the last unsuccessful segment of duration $\mathcal{S}$.  Thus the average lifetime of an effective forager is given by
\begin{align}
\label{T-well}
    \langle T\rangle \simeq \frac{Q_B}{1-Q}\;\tau_B(R)
    +\frac{Q_D}{1-Q} \;\tau_C+\mathcal{S}\,.
\end{align}
In Eqs.~\eqref{T-poor} and \eqref{T-well}, we make a mean-field approximation of replacing the average of the product of the number of segments multiplied by the time of each segment by the average number of segments multiplied by the average time of each segment.  As we discuss in the next section, the simulation data is is good agreement with the results of this naive approximation.

\section{Simulation and Empirics}

\begin{figure}[ht]
    \centering
    \includegraphics[width=0.7\textwidth]{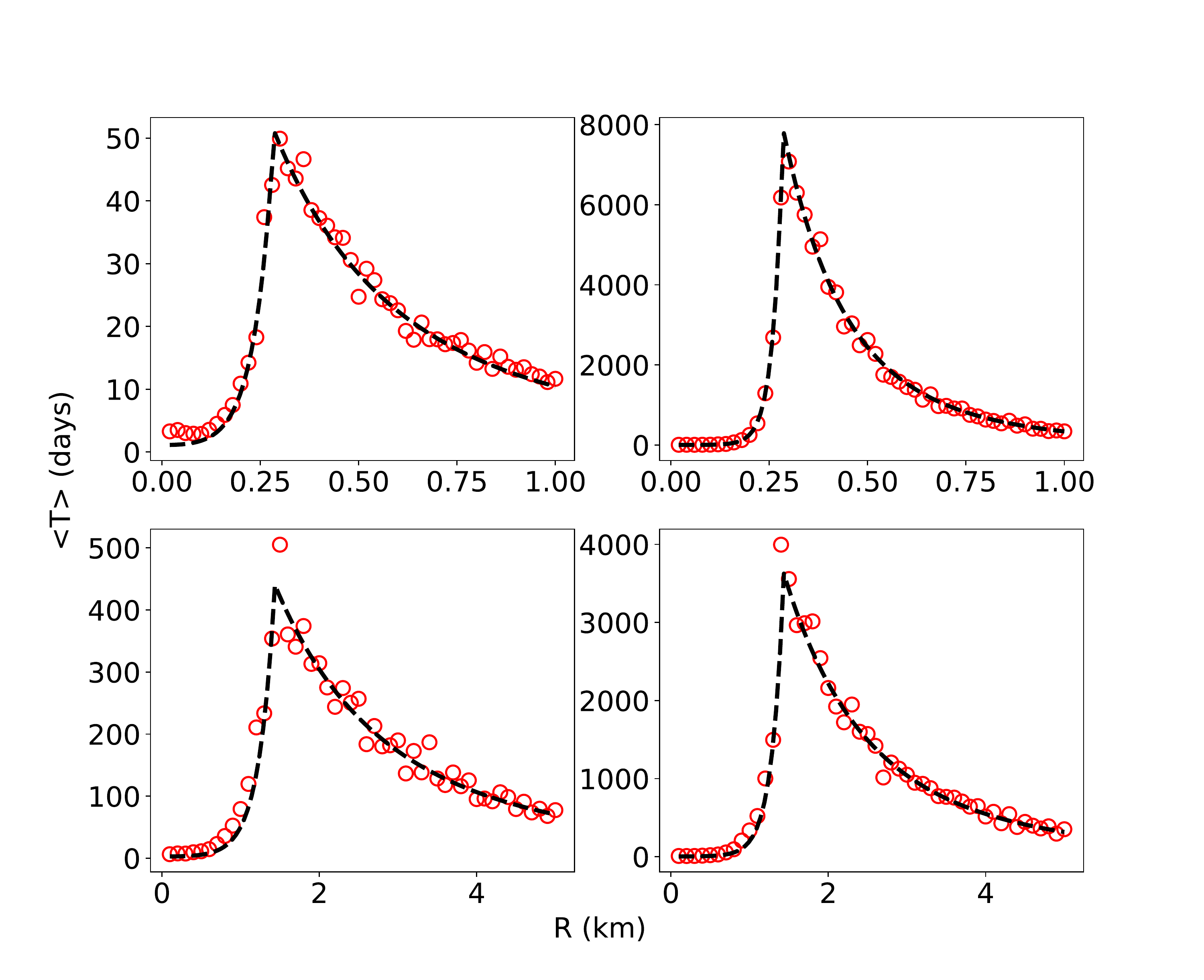}
    \caption{(top) Average lifetime $\left\langle T\right\rangle$ versus detection range $R$ for a $0.1$ kg forager in a one-dimensional environment, with average spacing between food $\ell = 50$ m (left) and $\ell=25$ m (right). (bottom) $\langle T\rangle$ versus $R$ for $1$ kg forager with $\ell = 0.2$ km (left), and $\ell = 0.15$ km (right). Circles represent simulation results, and dashed curves represent the analytical formula Eq.~\eqref{T-well}. }
    \label{fig:simulation_and_formula}
\end{figure}

\begin{figure}[ht]
\centerline{\includegraphics[width=1.\textwidth]{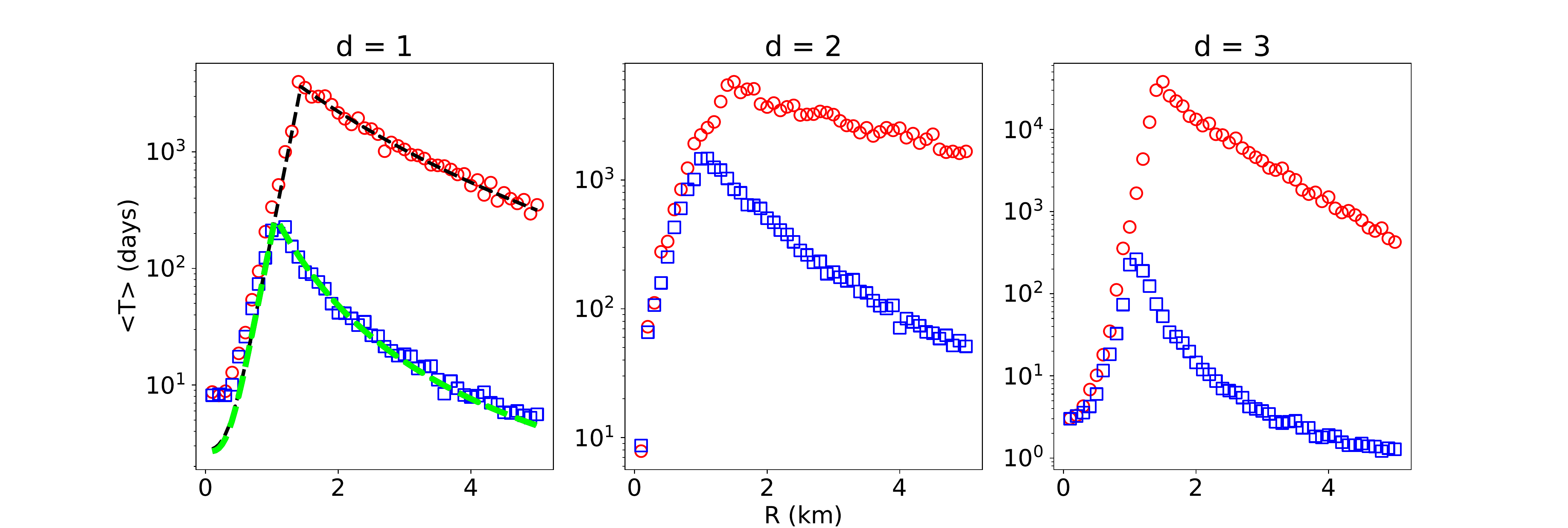}}
\caption{Average lifetime $\left\langle T\right\rangle$ versus detection range $R$ for a $1$ kg forager on a (left) one-dimensional, (middle) two-dimensional, and (right) three-dimensional environment, with average spacing between food $\ell = 0.15$ km (one dimension), $\ell = 0.2$ km (two dimension), $\ell = 0.8$ km (three dimensions). Red circles correspond to $\alpha = 0.1$ and blue squares correspond to $\alpha=0.5$.}
\label{fig:simulation_results}
\end{figure}

To check the analytical expressions for the forager lifetime in Eqs.~\eqref{T-poor} and \eqref{T-well}, we numerically simulate the fate of an intelligent forager in one, two and three dimensions.  It is first useful to fix the various model parameters in a way that has a connection to real-world foraging scenarios. We use the fact that the body mass of animals is strongly correlated with many physiological parameters~\cite{west1997general}, and we evaluate these physiological characteristics for a representative foraging organism of mass $0.1$ kg and of $1$ kg. We also use empirical and theoretical scaling laws to fix the forager's step size~\cite{carbone2005far}, basal metabolism~\cite{west1997general}, and food storage capacity~\cite{lindstedt2002use} within the framework of our intelligent forager model (see Table~\ref{tab:parameters_and_values}). We use these values in our simulations and calculations to explore the dependence of the forager lifetime on its detection range $R$, the density of food in the environment, and the detection-range dependence of its metabolic rate $\alpha$. We explore a range of the parameter $\alpha$ such that the additional intelligence-based contribution to a forager's metabolism is less than one-fifth~\cite{herculano2011scaling} and up to one-half  of the total metabolism (red circles in Figs.~\ref{fig:simulation_and_formula} \& \ref{fig:simulation_results} and blue squares in Fig.~\ref{fig:simulation_results}, respectively).

\begin{table}[ht]
    \centering
        \caption{Basic parameters and their corresponding numerical values for a forager of mass $0.1$ kg and $1$ kg in the intelligent forager model.}\medskip
    \begin{tabular}{|l|c|ccl|}
        \hline
        Definition & Parameter & \multicolumn{3}{c|}{Values and Units} \\
        \hline
        & & $m=0.1$ kg & $1$ kg &\\\cline{3-5}
        time step & $\delta t$ & $1$ & $1$ & day \\
        step size  \cite{carbone2005far} & $a \sim m^{0.3}$ & $0.3$ & $0.6\; $ & km\\
        metabolism without intelligence \cite{brown2004toward}& $B \sim m^{3/4}$ & $0.16$ & $0.9\;$ & MJ/day\\
        energy of full forager \cite{lindstedt2002use} & $E_0 \sim m^{1.14}$ & $0.18$ & $2.5\;$ & MJ\\
        metabolism range dependence \cite{herculano2011scaling} & $\alpha$ & $0.1$ & 0.1--0.5\; & MJ/day/km\\
        energy content of each food morsel & $F$ & $0.18$ & $2.5\;$ & MJ\\
        detection range & R & 2 & 10\; & km\\
        average distance between food & $\ell$ & 0.025--0.05 & 2.5--5\; & km\\
        \hline
    \end{tabular}
    \label{tab:parameters_and_values}
\end{table}

Our simulations are based on describing the forager motion as a combination of a continuous random walk, when the food is outside the forager's detection range, and directed motion towards the food when it is within the forager's detection range. To simulate infinitely large resource environments in a memory-efficient way, we create the resource landscape only in the vicinity of the forager's trajectory. Specifically, we specify the locations of the food within a given grid cell and its neighbors, only when the forager visits this cell for the first time. We set the size of the grid cell to be equal to the forager's detection range. 

The dependence of the forager lifetime on detection range $R$ in one dimension is shown in Fig.~\ref{fig:simulation_and_formula}. These simulation results and our analytical predictions of Eqs.~\eqref{T-poor} and \eqref{T-well} quantitatively agree in one dimension. We observe that the diffusive term in Eq.~\eqref{T-well} gives a small contribution to the overall lifetime compared to the ballistic term. We also see that for very small $R$, increasing the forager's intelligence is detrimental for its survival.  This effect is small and barely visible on the scale of Figs.~\ref{fig:simulation_and_formula} and \ref{fig:simulation_results}). This behavior arises because when $R\ll \ell$ the forager does not find food more efficiently than a dumb forager. However, the small intelligence that forager does possess imposes an additional metabolic cost. Consequently, the forger lifetime initially must decrease with $R$. Beyond a critical value of intelligence, however, the advantage in finding food more efficiently outweighs the additional metabolic cost of intelligence, so that the lifetime increases with intelligence.  Finally, for sufficiently large $R$, the additional metabolic cost of intelligence outweighs the benefit of a higher search efficiency, and the lifetime again decreases with $R$. The change from positive to negative slope in lifetime with respect to $R$ occurs when $R=v\mathcal{S}$ and corresponds to the transition from an effective to a feeble forager.

More importantly, the non-monotonic dependence of the forager lifetime on its detection range is robust phenomenon that arises in one, two, and three dimensions (Fig.~\ref{fig:simulation_results}).  The results in these figures show that there exists an optimal level of forager intelligence that maximizes its lifetime for the parameters given in Table~\ref{tab:parameters_and_values}. We also observe that the optimum detection range at which the lifetime is maximized depends weakly on $\alpha$ and $\ell$.

\section{Summary}

We introduced a model of intelligent foraging, in which intelligence is manifested by a forager being endowed with a detection range $R$ that is an increasing function of its intelligence.  Such an intelligent forager moves ballistically towards the nearest food that is within its detection range and wanders randomly otherwise.  A larger detection range makes the forager more efficient in finding food, but this larger range is accompanied by a larger metabolic cost for each step that the forager takes. Within a simple extension of the starving random walk model, we showed that this tradeoff gives rise to an optimum, intermediate level of intelligence that maximizes the lifetime of a forager.

This lifetime maximum is a robust phenomenon that we derived analytically in one dimension, and observed by simulations in one, two, and three dimensions.  While our modeling has ignored many important and realistic aspects of real forager behavior, incorporating them would make analytical modeling much more complicated. Within the framework of our intelligent forager model, it would be natural to assign a different metabolic cost to random wandering and to ballistic motion, as the latter may likely correspond to a forager running towards food.  It would also be appropriate to endow each food morsel with a different energy content so that consumption of a morsel of food may not necessarily lead to satiation of the forager.  In spite of the obvious limitations of our idealized foraging model, our approach seems to provide a useful starting point to understand the role of intelligence in foraging behavior. 

\section*{Acknowledgments}
We thank Justin Yeakel for stimulating discussions that helped nucleate this work. UB gratefully acknowledges financial support from NOAA's HPCC incubator grant, while SR gratefully acknowledges financial support from NSF grant DMR-1910736.



\providecommand{\newblock}{}

\end{document}